\documentclass[sn-basic]{sn-jnl}

\usepackage{graphicx}%
\usepackage{multirow}%
\usepackage{amsmath,amssymb,amsfonts}%
\usepackage{amsthm}%
\usepackage{mathrsfs}%
\usepackage[title]{appendix}%
\usepackage{xcolor}%
\usepackage{textcomp}%
\usepackage{manyfoot}%
\usepackage{booktabs}%
\usepackage{algorithm}%
\usepackage{algorithmicx}%
\usepackage{algpseudocode}%
\usepackage{listings}%

\usepackage{longtable}
\usepackage{amsmath}
\usepackage{multicol}
\RequirePackage{color}

\theoremstyle{thmstyleone}%
\theoremstyle{thmstyletwo}%

\theoremstyle{thmstylethree}%

\begin{document}

\title[Article Title]{A study on star formation in rotating and magnetized filamentary molecular clouds}


\author*[1,2]{\fnm{Ashok} \sur{Mondal}}\email{ashok.contai@gmail.com}

\affil*[1]{\orgdiv{Department of Applied Mathematics}, \orgname{University of Calcutta}, \orgaddress{\street{92 A.P.C Road}, \city{Kolkata}, \postcode{700009}, \state{West Bengal}, \country{India}}}

\affil[2]{\orgdiv{Department of Applied Science and Humanities}, \orgname{Haldia Institute of Technology}, \orgaddress{\street{Hatiberia, ICARE Complex}, \city{Haldia}, \postcode{721657}, \state{West Bengal}, \country{India}}}


\abstract{A 2D dynamic model is utilized to investigate star formation in rotating filamentary molecular clouds (FMCs) amidst magnetic fields. The study reveals that the emergence of field stars is possible under both weak and strong magnetic fields due to the presence of low-density structures. The presence of a strong rotation in a strongly magnetized FMC Plays a crucial role in forming a combination of two binary pairs and two field stars. An intermediate-density structure in the presence of a moderate magnetic field can form a combination of binary stars and field stars, whereas, in very high dense filamentary molecular clouds, a low magnetic field may help to form binary stars or stellar associations. In a rapidly rotating and weakly magnetized highly dense cloud binary stars are formed, whereas, strong rotation and moderate magnetic field help to form stellar associations.}




\maketitle

\section{Introduction}\label{sec1}

The interstellar medium (ISM) displays a predominantly filamentary structure across various scales, as evidenced by studies such as \citet{Schneider1979}, \citet{Hartmann2002}, \cite{Myers2009} and \citet{Flagey2009}. These self-gravitating filaments are generally found in dense molecular clouds that serve as the birthplaces of stars.  Star-forming molecular clouds are threaded by magnetic fields, which certainly influence cloud morphology and evolution (\citealt{Krumholz_2019}). In Galactic molecular clouds, magnetic fields are frozen-in fields from the interstellar medium of the Milky Way, which arose from galactic-scale dynamo amplification of seed fields generated by Biermann batteries in early Active Galactic Nuclei and/or in Population III stars. Currently, the study of magnetic fields in molecular clouds is an important and challenging topic in itself to know the role of magnetic fields in the formation and evolution of interstellar clouds and in star formation.
Despite significant progress in recent years, the basic physics of star formation is not clearly explained. One of the most fundamental questions about star formation that has remained unanswered is what drives the star formation process. Depending on the role of magnetic fields there are two major classes of star formation theory: the strong-field models and the weak-field models. In strong magnetic field models, magnetic fields with ambipolar diffusion control the formation and evolution of the molecular clouds from which stars form. Ambipolar diffusion drives to form dense cores and their gravitational collapse to form protostars. In this theory, clouds are formed with subcritical masses, and hence, the magnetic pressure is sufficiently strong to counteract gravity and prevent gravitational collapse. 
Neutral gases and dust contract gravitationally through the magnetic field and ions because the magnetic field is only frozen into the ionized gas and dust, increasing mass in the cloud cores. The magnetic field strength increases more slowly than does mass.  In weak magnetic field models, turbulent flows control the formation of clouds and dense cores, with cores either dissipating back into the general interstellar medium or collapsing and forming stars if they are self-gravitating when formed. In this theory, clouds are formed with supercritical masses, hence in general, clouds dissipate as they do not become gravitationally bound; those that are self-gravitating form stars in essentially a free-fall time.
This study delves into the evolution of a collection of nonlinear dynamical equations, derived from an extended form of the double-well potential. I explore diverse physical scenarios to unravel the mechanisms giving birth to field, binary, and multiple star systems in a rotating and  magnetized molecular clouds. \\
\noindent
In Section \ref{mathematical model}, the mathematical model is expounded upon. Following this, Section \ref{sec:initial} outlines and deliberates upon the initial parameter values. Section \ref{results and discussion} encompasses the presentation and discourse of results, while the concluding remarks can be found in Section \ref{sec:conclusions}. Additional mathematical derivations are provided in Appendix \ref{appendix}.

\section{Mathematical Model}
\label{mathematical model}
In this work, I have studied the dynamical evolution of filamentary molecular clouds (hereafter FMCs) in the presence of the magnetic field. \citet{Mondal_2021} studied the evolution of rotating FMCs. For this purpose, they considered double-well potential (hereafter DWP), which has similar density structures to FMCs. They introduced the following generalized DWP (Eq.\ref{eq:GDWP}) in a two-dimensional system and compared the similarity between DWP and FMCs.
\begin{equation}
V(X,Y,Z^2_{x},Z^2_{y})=-Z^2_{x}X^2-Z^2_{y}Y^2+\lambda(a_{xx}X^4+2a_{x y}X^2Y^2+a_{y y}Y^4).
   \label{eq:GDWP}
\end{equation}
Details of these parameters $Z_x^2$, $Z_y^2$, $a_{xx}$, $a_{xy}$, $a_{yy}$, and $\lambda$, and their significance can be found in \citet{Mondal_2021}.
In this work, I further study the evolution in presence of a constant magnetic field first, and then the evolution of rotating and magnetized FMCs.

\subsection{Presence of a constant magnetic field}
\label{sec:magnetic field}
The role of the magnetic field on the formation and evolution of FMCs have been studied by many researchers through theoretical and observational investigations ( \citealt{Khesali2014}; \citealt{Kainulainen2016}; \cite{Aghili2017}; \citealt{Tokuda2019}). This field firmly impacts the shape of interstellar gas through the formation of filaments and reduction of the numbers of clumps, cores, and stars (\citet{inutsuka_2018}; \citet{Hennebelle2019}). Self- gravitating FMCs consistently attempt to get steady against fragmentation while the magnetic field influences interstellar gas motion for fragmentation. Numerous FMCs are associated with magnetic fields having energy comparable to the gravitational binding energy(\citealt{Myers1988}; \citealt{Carlberg1990}; \citealt{Li2016}; \citealt{Klassen2016}). In our model, I have considered the presence of a constant magnetic field in the filamentary molecular cloud.
\noindent
The direction of the magnetic field in the spiral galaxy is along the direction of spiral arms (\citealt{Sofue_1986}, \citealt{Beck_2015}). So it has no component along z-direction, i.e, $ \Vec{B} = (B_{x},B_{y},0)$, and the corresponding magnetic force components along x and y direction are 
      $- \frac{B^{2} x}{x^{2}+y^{2}}$ and $- \frac{B^{2} y}{x^{2}+y^{2}}$ respectively. So in the presence of a constant magnetic field  strength B along the spiral arm, the scaled equations (See Appendix \ref{appendix}) of motion for general DWP are,

\begin{equation}
  \ddot x = 2 Z^2_{x}x -\lambda(4a_{xx}x^3+4a_{x y}x y^2)-\frac{1}{\rho}\frac{B^2 x}{x^2+y^2},
  \label{eq:GMX}
    \end{equation}
\begin{equation}
  \ddot y = 2 Z^2_{y}y -\lambda(4a_{y y}y^3+4a_{x y}x^2 y)-\frac{1}{\rho}\frac{B^2 y}{x^2+y^2},
  \label{eq:GMY}
    \end{equation}
where $\rho$ is the density of the molecular clouds. Here, I have made dimensionless (See Appendix \ref{appendix}) X, Y, and T by $X=x\times10^{17}$ cm, $Y=y\times10^{17}$ cm, and $T=t\times 10^{13}$ s for observational compatibility. \\ 
\noindent
For stability analysis, I transformed the aforementioned second-order differential equations into a system of first-order differential equations. Then equations \ref{eq:GMX} and \ref{eq:GMY} reduce to the following system of equations.\\
\noindent $ \dot x=X'$\\
$ \dot X' = 2 Z^2_{x}x -\lambda(4a_{xx}x^3+4a_{x y}x y^2) -\frac{1}{\rho}\frac{B^2 x}{x^2+y^2}$\\
$ \dot y= Y'$\\
$ \dot Y' = 2 Z^2_{y}y -\lambda(4a_{y y}y^3+4a_{x y}x^2 y) -\frac{1}{\rho}\frac{B^2 y}{x^2+y^2}$.\\
\noindent
The number of solutions or fixed points are 16 of which 8 of them can be written as\\
     (i) $\{x=0, y= \pm \frac{ \sqrt{ \rho Z^2_{y} \pm \sqrt{ (\rho Z^2_{y})^2 - 4  \lambda \rho B^2 a_{y y}}} }{2 \sqrt{ \rho \lambda a_{y y}}}\},$\\
     (ii) $\{ x= \pm \frac{ \sqrt{ \rho Z^2_{x} \pm \sqrt{ (\rho Z^2_{x})^2 - 4  \lambda \rho B^2 a_{x x}}} }{2 \sqrt{ \rho \lambda a_{x x}}}, y=0 \},$\\
     The mathematical expression for the remaining stationary points are very large to be quoted. I have studied the stability of this system of stationary points in Table~\ref{TAble_potential_magnetic} by considering suitable arbitrary values of the parameters $ \lambda$, $ a_{x x}$, $a_{x y}$, $a_{y y}$, $Z^2_{x}$ and $Z^2_{y}$, for which the eigenvalues are real negative or their real parts are close to zeros as found through numerical computations. The eigenvalues of the corresponding Jacobian matrix can be expressed as,  $\pm \frac{\sqrt{(P_{2}+R_{2}) \pm \sqrt{(P_{2}-R_{2})^2 + 4Q^2_{2}}}}{\sqrt{2}}$ where, $P_{2}= 2Z^2_{x} - 12 \lambda a_{x x} x^2 - 4 \lambda a_{x y} y^2 + \frac{B^2(x^2 - y^2)}{ \rho (x^2 + y^2)^2}$, $ Q_{2} = -8 \lambda a_{x y} x y + \frac{2 B^2 x y}{ \rho (x^2 + y^2)^2}$ and  $R_{2}= 2Z^2_{y} - 12 \lambda a_{y y} y^2 - 4 \lambda a_{x y} x^2 - \frac{B^2(x^2 - y^2)}{ \rho (x^2 + y^2)^2}$. One sufficient condition for which a stationary point becomes a stable center is $ P_{2} R_{2} > Q^2_{2}$. In stable center scenario, particles near a stable point will move around that stable point, and this situation might attract other points to come closer, which helps later to collapse towards the stable point. I have not studied chaos as it has no relevance in the present study for star formation.   

    \begin{figure}[t]
	\includegraphics[width=\columnwidth]{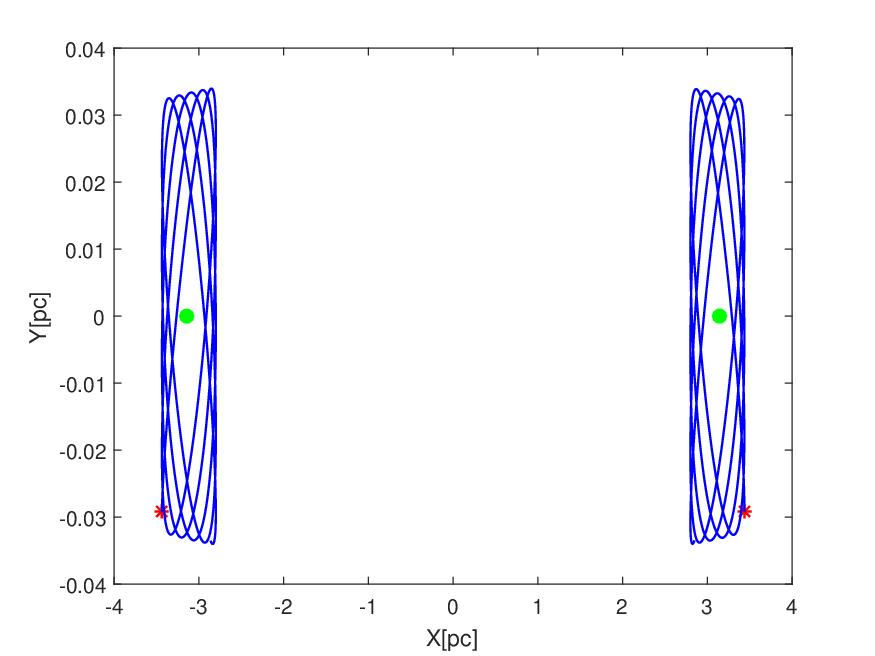}
    \caption{Phase space for the gravitational potential in the presence of a constant magnetic field $B =10 \mu G$ without any rotation for the values of the parameters $\lambda=0.25\times 10^{-5}, a_{x x}= 1, a_{x y}=1, a_{y y}=2, Z^2_{x}= 0.05, Z^2_{y}=-0.05.$. The stable center stationary points are marked by green dots,The blue curves are particular solutions starting at the positions indicated by red stars.}
    \label{fig: magnetic filed}
\end{figure}

\subsection{Presence of both 
rotation and magnetic field}
In this section I discuss the fragmentation and stability of FMCs rotating with a constant angular velocity in the presence of a constant magnetic field throughout the cloud.
The scaled equations(See Appendix \ref{appendix}) of motion for the general DWP are
\begin{equation}
  \ddot x = 2 Z^2_{x}x -\lambda(4a_{xx}x^3+4a_{x y}x y^2)-\frac{1}{\rho}\frac{B^2 x}{x^2+y^2} +\Omega^2 x +2\Omega \dot y
  \label{eq:GMRX}
    \end{equation}
\begin{equation}
  \ddot y = 2 Z^2_{y}y -\lambda(4a_{y y}y^3+4a_{x y}x^2 y)-\frac{1}{\rho}\frac{B^2 y}{x^2+y^2} + \Omega^2 y - 2 \Omega \dot x.
  \label{eq:GMRY}
    \end{equation}
\noindent
I have converted this system of second order differential equations to a system of first order differential equations and proceeded as in Section ~\ref{sec:magnetic field} for stability analysis. The number of solutions or fixed points are 16 out of which 8 of them can be written as, \\
     (i) $\{x=0, y= \pm \frac{ \sqrt{ \rho (2 Z^2_{y} + \Omega^2) \pm \sqrt{ (\rho^2(2 Z^2_{y} + \Omega^2)^2 - 16  \lambda \rho B^2 a_{y y}}} }{2 \sqrt{ 2\rho \lambda a_{y y}}}\},$\\
     (ii) $\{ x= \pm \frac{ \sqrt{ \rho (2 Z^2_{x} + \Omega^2) \pm \sqrt{ (\rho^2(2 Z^2_{x} + \Omega^2)^2 - 16  \lambda \rho B^2 a_{x x}}} }{2 \sqrt{ 2\rho \lambda a_{x x}}} , y=0 \},$\\
     The mathematical expression for remaining stationary points are large to be quoted. I have studied the stability for the system of stationary points in Table~\ref{TAble_potential_magnetic_rotation} by considering suitable arbitrary values of the parameters $ \lambda$, $ a_{x x}$, $a_{x y}$, $a_{y y}$, $Z^2_{x}$ and $Z^2_{y}$ for which the eigenvalues have real values or real parts are close to zero. The eigenvalues of the corresponding Jacobian matrix can be expressed as  $\pm \frac{\sqrt{(P_{4}+R_{4}-4\Omega^2) \pm \sqrt{(P_{4}+R_{4} -4\Omega^2 )^2 - 4(P_{4}R_{4}-Q^2_{4}))}}}{\sqrt{2}}$,\\ where, $P_{4}= 2Z^2_{x} - 12 \lambda a_{x x} x^2 - 4 \lambda a_{x y} y^2 + \frac{B^2(x^2 - y^2)}{ \rho (x^2 + y^2)^2} + \Omega^2$, $ Q_{4} = -8 \lambda a_{x y} x y + \frac{2 B^2 x y}{ \rho (x^2 + y^2)^2}$ and  $R_{4}= 2Z^2_{y} - 12 \lambda a_{y y} y^2 - 4 \lambda a_{x y} x^2 - \frac{B^2(x^2 - y^2)}{ \rho (x^2 + y^2)^2} + \Omega^2$.\\
     One sufficient condition for which a stationary point becomes a stable center is,  $ P_{4} R_{4} > Q^2_{4}.$

\section{Initial values of the parameters}
\label{sec:initial} 
I have studied the stability of different types of molecular clouds (MCs) for a wide range of the  parameters. \citet{Khesali2014} showed that the magnetic field in the MCs is of the order $\mu$G. The range of mean magnetic field strength in giant molecular cloud (GMC) is from  $ 1.4 \, \mu G$ to $ 14 \, \mu G $ \citep{Ostriker2001}. In the present work, I have considered the range of B from $ 1.4 \, \mu G$ to  $14 \, \mu G$. Several authors (\citet{Herbst1973}; \citet{ Goldsmith1978}; \citet{Bally1987, Bally1988}; \citealt{Caselli1999}) found the initial number density of MCs as $\sim$ $10^4$ cm$^{-3}$ to $ 10^5$ cm$^{-3} $, and the peak density as $\sim$ $ 10^5 $ cm$^{-3} $ \citep{Myers2017}. Although some authors (\citealt{Gammie1996}; \citealt{Ostriker1999})) showed the mean cloud density $n_{H_{2}}$ is typically $\sim$ 25 cm$^{-3}$ to 100 cm$^{-3}$ in cloud complexes, whereas in clumps it is  $ 10^3 $ cm$^{-3}$. Throughout the current work, I have considered the initial number density in MCs as $ 10^4$ cm$^{-3}$ i.e, $ \rho = 1.67 \times 10^{-20}$ g\,cm$^{-3} $. 
\citet{Brainel2018} shown the angular velocity of most of the MCs is less than $ 0.03 $\,km\,s$^{-1}$\,pc$^{-1}$ \citep{Brainel2018}, although it may be as high as $ 1.5$\,km\,s$^{-1}$\,pc$^{-1}$ in few cases \citep{Goldreich1974}. The angular velocity of different types of cores in a MC is higher than the average angular velocity of that MC. The observed value of $ \Omega$ of cloud cores is in the range from 0.3 \,km\,s$^{-1}$\,pc$^{-1}$ to 3 \,km\,s$^{-1}$\,pc$^{-1}$ (\citealt{Goodman1993}; \citealt{Barranco1998}; \citealt{Burkert2000}). In this work, I have considered the full range of rotational velocity to understand full effect of rotation i.e. $\Omega$ varies from 0.01 \,km\,s$^{-1}$\,pc$^{-1}$  to 3 \,km\,s$^{-1}$\,pc$^{-1}$.
\section{Results and discussion}
\label{results and discussion}
\begin{figure}
	\includegraphics[width=\columnwidth]{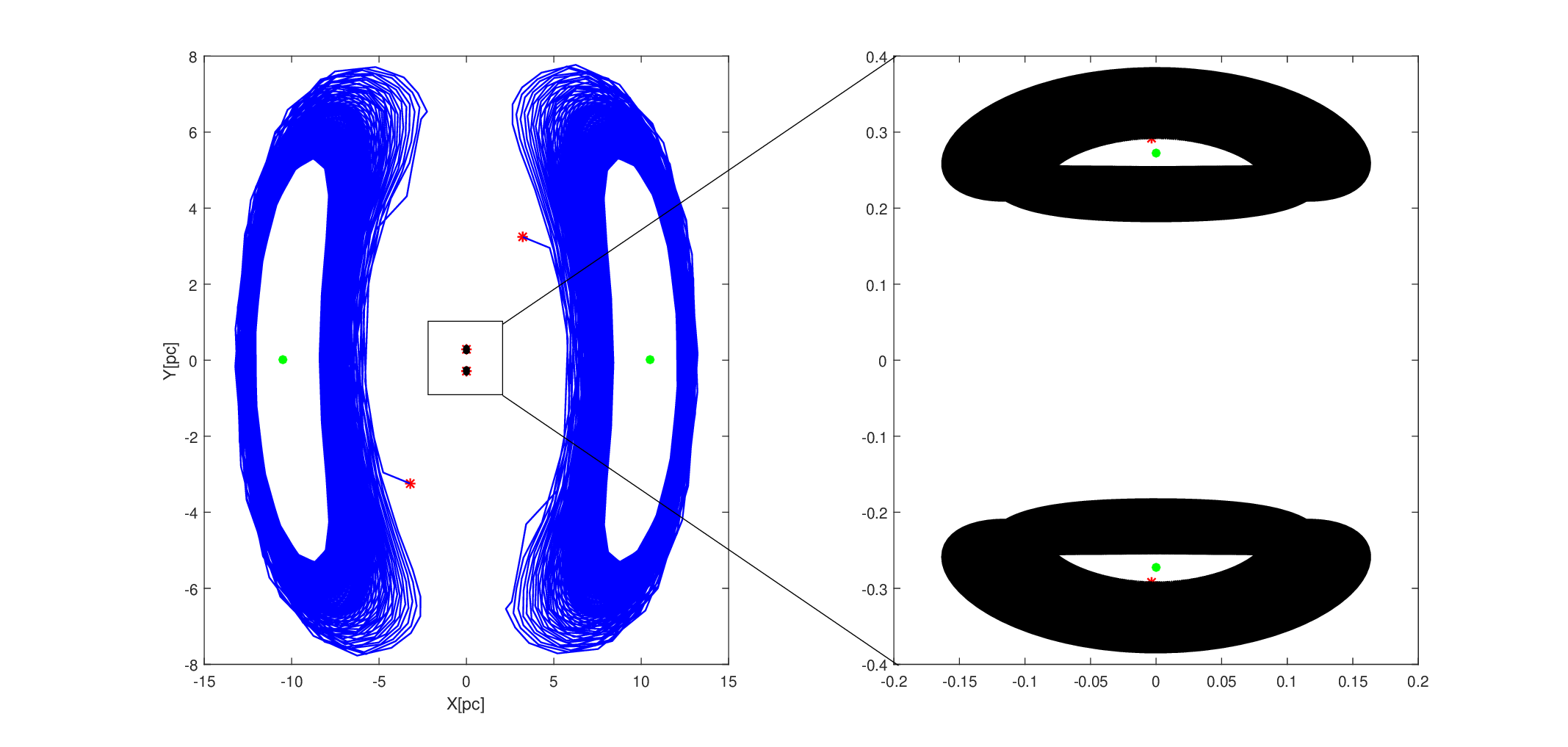}
    \caption{Phase space for a self-gravitating and magnetized rotating cloud with $\Omega = 3\times 10^{-13}s^{-1}$, and $B = 10 \mu G$ as in Fig. \ref{fig: magnetic filed} for the same values of the parameters. For these values one binary pairs and two isolated stars are formed. }
    \label{fig: magnetic field rotation}
\end{figure}

In this study, I have developed a synthetic model to investigate the potential of DWP to facilitate the formation of field stars, binary pairs, or stellar associations within FMCs. \citet{Mondal_2021} showed that the density patterns associated with a DWP can exhibit resemblances to those observed in FMCs , which are the remnants of strong shock waves passing through them in the presence of a helical magnetic field ( \citealt{Pudritz2013}; \citealt{Fiege2000}; \citealt{McKee1993}; \citealt{Heiles1993}; \citealt{Schleuning1998}; \citealt{Mondal2019}). I have explored the existence of stable stationary points for various values of the parameters in the DWP along with various values of the strength of the magnetic field ($B$) and rotational velocity ($\Omega$). The results have been shown in Tables \ref{TAble_potential_magnetic} - \ref{TAble_potential_magnetic_rotation} and Figs. \ref{fig: magnetic filed} - \ref{fig: magnetic field rotation}. 
Figs. \ref{fig: magnetic filed} and \ref{fig: magnetic field rotation} show that the particles very close to a stable center point come around again and again. I get ellipsis in the phase plane, and consequently, particles of the clouds will be exaggerated around the stable points. Thus, mass around a stable center increases, and subsequently, when it crosses its jeans mass, collapse starts, and the star formation process begins. Drawing from \citet{Jiménez-Esteban_2019} analysis of 3741 commoving binary and multiple stellar systems in the Gaia DR2 catalogue, \citet{Mondal_2021} have considered the maximum distance between two binary stars as 2.45pc.
Table \ref{TAble_potential_magnetic} compiles the results in the presence of magnetic field. The following facts are observed from Table \ref{TAble_potential_magnetic}, which is obtained for magnetized self-gravitating FMC.\\

\noindent
When $\lambda$ is very small ($\sim 10^{-5}$ ) and DWP is asymmetric (here, $a_{xx}=a_{xy}=1,a_{yy}=2, Z_x^2=0.05$ and $Z_y^2=-0.05$) or when $\lambda$ is higher ($\sim 1 $ ) and the remaining parameters ($a_{xx}=a_{xy}=10^{-7},a_{yy}=2\times10^{-7}, Z_x^2=10^{-2}$ and $Z_y^2=1.96\times10^{-2}$) are small, then field star formation is very likely. 
These values of the parameters indicate a small depth of the DWP, i.e., the density distribution of the DWP is lighter, which corresponds to scattered FMC (\citealt{Mondal_2021}). In these self-gravitating FMCs, a higher magnetic field only reduces the distance between two stable points. Thus, when B increases the distance between stable points decreases in lower dense FMC.

\noindent
When the density of an FMC is moderate ( $\lambda \sim 10^{-3}$ and $a_{xx}=a_{xy}=1,a_{yy}=2, Z_x^2= 1$ and $Z_y^2=1.96$), then in the presence of a higher magnetic field ( $B \sim 10 \mu G$  ) binary pairs form, whereas in the presence of a very low magnetic field ($B \sim 1.4 \mu G$ ) stellar associations are formed. Since $\lambda$ is smaller and the remaining parameters are higher compared to previous values, their combined effect indicates a moderate depth of the DWP i.e., moderately dense FMC.

\noindent
In a highly dense FMC ($\lambda \sim 10^{-2}$ and $a_{xx}=a_{xy}=1,a_{yy}=2, Z_x^2= 1$ and $Z_y^2=1.96$ ) stars may or may not form depending on the strength of the magnetic field. In presence of a lower magnetic field binary pairs or stellar associations are formed, whereas in the presence of a strong magnetic field no stars will form i.e. there are no stable centers. In high-dense FMCs, the strict magnetic flux freezing can not be assumed to hold always, especially when the ambipolar diffusion time scale becomes much shorter. Then, the neutrals in a magnetized cloud drift past the ions and the magnetic field lines, and consequently the radius of the FMC decreases. As we have considered the presence of a constant magnetic field, the magnetic critical mass decreases, so, FMCs get a favour to collapse and fragment into smaller parts. These fragments may undergo further subsequent fragmentation due to their low magnetic critical mass, and hence very low-mass fragments may be formed. The fragmentation of molecular cloud can give rise to gaseous fragments with masses 0.001 $M_\odot$ (\citealt{Mondal_ASHOK_2019}, \citealt{Whitworth_2006}, \citealt{Boss_2001}), but The lowest critical mass for a star to support nuclear fusion is 0.08 $M_\odot$ (\citealt{Nakamura_1997}). Thus, highly magnetized and highly dense FMCs sometimes may remain starless. 

\setlength{\tabcolsep}{1.2pt} 
\renewcommand{\arraystretch}{1.0}
\begin{longtable}{ c c c c c c c c c c}
\caption{Number of stationary points and stable centers for various types of self-gravitating magnetized filamentary MC.}
\label{TAble_potential_magnetic}\\ \hline
 & & & & & & &  No. of & No. of \\
 \multicolumn{1}{c}{\textbf{$\lambda$}}  & \multicolumn{1}{c}{\textbf{$a_{x x}$}} & \multicolumn{1}{c}{\textbf{$a_{x y}$}} & \multicolumn{1}{c}{\textbf{$a_{y y}$ }}& \multicolumn{1}{c}{\textbf{$Z^2_{x}$ }}& \multicolumn{1}{c}{\textbf{$ Z^2_{y}$}} & \multicolumn{1}{c}{\textbf{$B$}}& \multicolumn{1}{c}{stationary} & \multicolumn{1}{c}{stable centers}\\
 & & & & & &  ($\mu$G) & points & (X, Y):distance (pc)\\
& & & & & &   & &   (pc, pc)\\

\endfirsthead
\multicolumn{3}{c}%
{{\bfseries \tablename\ \thetable{}-continued}} \\
 \hline
 & & & & & & &  No. of & No. of \\
 \multicolumn{1}{c}{\textbf{$\lambda$}}  & \multicolumn{1}{c}{\textbf{$a_{x x}$}} & \multicolumn{1}{c}{\textbf{$a_{x y}$}} & \multicolumn{1}{c}{\textbf{$a_{y y}$ }}& \multicolumn{1}{c}{\textbf{$Z^2_{x}$ }}& \multicolumn{1}{c}{\textbf{$ Z^2_{y}$}} & \multicolumn{1}{c}{\textbf{$B$}}& \multicolumn{1}{c}{stationary} & \multicolumn{1}{c}{stable centers}\\
 & & & & & &  ($\mu$G) & points & (X, Y):distance (pc)\\
& & & & & &   & &   (pc, pc)\\
\endhead
\hline
\endfoot
\hline
$0.25\mathrm{e}{-5}$ & 1 & 1 & 2 & 0.05 & -0.05& 10 & 4 & 2 \\
& & & & & &  & & $( \pm 3.135, 0)$ \\

$0.25\mathrm{e}{-5}$ & 1 & 1 & 2 & 0.05 & -0.05& 14 & 4 & 2 \\
& & & & & &  & & $( \pm 3.012, 0)$ \\

$0.25\mathrm{e}{-5}$ & 1 & 1 & 2 & 0.05 & -0.05& 5 & 4 & 2 \\
& & & & & &  & & $( \pm 3.215, 0)$ \\

$0.25\mathrm{e}{-5}$ & 1 & 1 & 2 & 0.5 & -0.005& 10 & 4 & 2 \\
& & & & & &  & & $( \pm 32.404, 0)$ \\

$0.25\mathrm{e}{-5}$ & 1 & 1 & 2 & 0.6 & 0.5& 10 & 4 & 2 \\
& & & & & &  & & $( \pm 11.223, 0)$ \\

$0.25\mathrm{e}{-5}$ & 1 & 1 & 2 & 0.5 & 0.6 & 10 & 12 & 4 \\
& & & & & &  & & $( \pm 9.162, \pm 4.583)$ \\

$0.25\mathrm{e}{-4}$ & 1 & 1 & 2 & $0.05\mathrm{e}{+1}$ & $-0.05\mathrm{e}{+1}$ & 6 & 4 & 2 \\
& & & & & &  & & $( \pm 3.237, 0)$ \\

$0.25\mathrm{e}{-4}$ & 1 & 1 & 2 & $0.05\mathrm{e}{+1}$ & $-0.05\mathrm{e}{+1}$ & 1.4 & 4 & 2 \\
& & & & & &  & & $( \pm 10.247, 0)$ \\

$1$ & $1\mathrm{e}{-7}$ &  $1\mathrm{e}{-7}$  &  $2\mathrm{e}{-7}$  & $1\mathrm{e}{-2}$ & $1.96\mathrm{e}{-2}$ & 10 & 8 & 2 \\
& & & & & &  & & $(0, \pm 7.057)$ \\

$1$ & $1\mathrm{e}{-7}$ &  $1\mathrm{e}{-7}$  &  $2\mathrm{e}{-7}$  & $1\mathrm{e}{-2}$ & $1.96\mathrm{e}{-2}$ & 5 & 12 & 4 \\
& & & & & &  & & $(\pm 1.141, \pm 7.099)$ \\

$1$ & $1\mathrm{e}{-7}$ &  $1\mathrm{e}{-7}$  &  $2\mathrm{e}{-7}$  & $1\mathrm{e}{-2}$ & $1.96\mathrm{e}{-2}$ & 14 & 8 & 2 \\
& & & & & &  & & $(0, \pm 6.935 )$ \\

$2$ & $1\mathrm{e}{-7}$ &  $1\mathrm{e}{-7}$  &  $2\mathrm{e}{-7}$  & $1\mathrm{e}{-2}$ & $1.96\mathrm{e}{-2}$ & 10 & 8 & 2 \\
& & & & & &  & & $(0, \pm 4.900)$ \\

$1$ & $1\mathrm{e}{-7}$ &  $1\mathrm{e}{-7}$  &  $2\mathrm{e}{-7}$  & $1\mathrm{e}{-1}$ & $1.96\mathrm{e}{-1}$ & 10 & 12 & 4 \\
& & & & & &  & & $(\pm 4.548, \pm 22.450)$ \\
0.25 & $1\mathrm{e}{-7}$ & 0 & 0 & $0.5\mathrm{e}{-2}$ & $-0.5\mathrm{e}{-2}$ & 10 & 4 & 2 \\
& & & & & &  & & $( \pm 9.912, 0)$ \\

0.25 & $1\mathrm{e}{-7}$ & 0 & 0 & $0.5\mathrm{e}{-2}$ & $-0.5\mathrm{e}{-2}$ & 5 & 4 & 2 \\
& & & & & &  & & $( \pm 10.167, 0)$ \\

0.25 & $1\mathrm{e}{-7}$ & 0 & 0 & $0.5\mathrm{e}{-2}$ & $-0.5\mathrm{e}{-2}$ & 14 & 4 & 2 \\
& & & & & &  & & $( \pm 9.524, 0)$ \\

$0.25\mathrm{e}{-1}$ & $1\mathrm{e}{-7}$ & 0 & 0 & $0.5\mathrm{e}{-2}$ & $-0.5\mathrm{e}{-2}$ & 10 & 4 & 2 \\
& & & & & &  & & $( \pm 32.301, 0)$ \\

0.25 & $1\mathrm{e}{-7}$ & 0 & 0 & 0.5 & -0.5 & 10 & 4 & 2 \\
& & & & & &  & & $( \pm 102.455, 0)$ \\

$1\mathrm{e}{-3}$ & 1 & 1 & 2 & 1 & 1.96 & 10 & 8 & 2 \\
& & & & & &  & & $( 0, \pm 0.7057)$ \\

$1\mathrm{e}{-3}$ & 1 & 1 & 2 & 1 & 1.96 & 1.4 & 12 & 4 \\
& & & & & &  & & $( \pm 0.1428, \pm 0.7099)$ \\

$1\mathrm{e}{-3}$ & 100 & 100 & 200 & 1 & 1.96 & 10 & 0 & 0 \\

$1\mathrm{e}{-3}$ & 100 & 100 & 200 & 1 & 1.96 & 1.4 & 8 & 2 \\
& & & & & &  & & $( 0, \pm 0.0693)$ \\
$0.25\mathrm{e}{-4}$ & 1 & 1 & 2 & 0.05 & -0.05& 10 & 0 & 0 \\

$0.25\mathrm{e}{-4}$ & 1 & 1 & 2 & 0.05 & -0.05& 6 & 4 & 2 \\
& & & & & &  & & $( \pm 0.8484, 0)$ \\

$0.25\mathrm{e}{-5}$ & 1000 & 1000 & 2000 & 0.05 & -0.05 & 10 & 0 & 0 \\

$0.25\mathrm{e}{-5}$ & 1000 & 1000 & 2000 & 0.05 & -0.05 & 1.4 & 0 & 0 \\

$1\mathrm{e}{-2}$ & 1 & 1 & 2 & 1 & 1.96 & 10 & 0 & 0 \\

$1\mathrm{e}{-2}$ & 1 & 1 & 2 & 1 & 1.96 & 14 & 0 & 0 \\

$1\mathrm{e}{-2}$ & 1 & 1 & 2 & 1 & 1.96 & 1.4 & 12 & 4 \\
& & & & & &  & & $( \pm 0.0382, \pm 0.2242)$ \\

$1\mathrm{e}{-2}$ & 1 & 1 & 2 & 1 & 1.96 & 5 & 8 & 2 \\
& & & & & &  & & $( 0, \pm 0.2168)$ \\

\hline 
\end{longtable}

\noindent

Table \ref{TAble_potential_magnetic_rotation} compiles the results of the analysis in the presence of both rotation and magnetic field. 
The following facts have been observed.

\begin{footnotesize}
\setlength{\tabcolsep}{1pt} 
\renewcommand{\arraystretch}{1.0}
\begin{longtable}{ c c c c c c c c c c c}
\caption{Number of stationary points and stable centers for various types of self-gravitating, magnetized, and rotating filamentary MC.}
\label{TAble_potential_magnetic_rotation}\\ \hline
 & & & & & & & & No. of & No. of \\
 \multicolumn{1}{c}{\textbf{$\lambda$}}  & \multicolumn{1}{c}{\textbf{$a_{x x}$}} & \multicolumn{1}{c}{\textbf{$a_{x y}$}} & \multicolumn{1}{c}{\textbf{$a_{y y}$ }}& \multicolumn{1}{c}{\textbf{$Z^2_{x}$ }}& \multicolumn{1}{c}{\textbf{$ Z^2_{y}$}} & \multicolumn{1}{c}{\textbf{$\Omega$}} & \multicolumn{1}{c}{\textbf{$B$}}& \multicolumn{1}{c}{stationary} & \multicolumn{1}{c}{stable centers}\\
 & & & & & & (km\,s$^{-1}$\,pc$^{-1})$&  ($\mu$G) & points & (X, Y):distance (pc)\\
& & & & & & &  & &   (pc, pc)\\

\endfirsthead
\multicolumn{3}{c}%
{{\bfseries \tablename\ \thetable{}-continued}} \\
 \hline
 & & & & & & & & No. of & No. of \\
 \multicolumn{1}{c}{\textbf{$\lambda$}}  & \multicolumn{1}{c}{\textbf{$a_{x x}$}} & \multicolumn{1}{c}{\textbf{$a_{x y}$}} & \multicolumn{1}{c}{\textbf{$a_{y y}$ }}& \multicolumn{1}{c}{\textbf{$Z^2_{x}$ }}& \multicolumn{1}{c}{\textbf{$ Z^2_{y}$}} & \multicolumn{1}{c}{\textbf{$\Omega$}} & \multicolumn{1}{c}{\textbf{$B$}}& \multicolumn{1}{c}{stationary} & \multicolumn{1}{c}{stable centers}\\
 & & & & & & (km\,s$^{-1}$\,pc$^{-1})$&  ($\mu$G) & points & (X, Y):distance (pc)\\
& & & & & & &  & &   (pc, pc)\\
\endhead
\hline
\endfoot
\hline

$0.25\mathrm{e}{-5}$ & 1 & 1 & 2 & 0.05 & -0.05 & 3 & 10 & 8 & 4 \\
& & & & & &  & & & $( 0, \pm 0.2730)$, $ (\pm 10.474,0)$ \\

$0.25\mathrm{e}{-5}$ & 1 & 1 & 2 & 0.05 & -0.05 & 3 & 14 & 8 & 4 \\
& & & & & &  & & & $( 0, \pm 0.3824)$, $ (\pm 10.471,0)$ \\

$0.25\mathrm{e}{-5}$ & 1 & 1 & 2 & 0.05 & -0.05 & 3 & 5 & 8 & 4 \\
& & & & & &  & & & $( 0, \pm 0.1361)$, $ (\pm 10.476,0)$ \\

$0.25\mathrm{e}{-5}$ & 1 & 1 & 2 & 0.05 & -0.05 & 0.01 & 10 & 4 & 2 \\ & & & & & &  & & & $ (\pm 3.216,0)$ \\

$0.25\mathrm{e}{-5}$ & 1 & 1 & 2 & 0.05 & -0.05 & 1.5 & 10 & 8 & 4 \\ & & & & & &  & & & $( 0, \pm 0.3422)$, $ (\pm 5.939,0)$ \\

$0.25\mathrm{e}{-5}$ & 1 & 1 & 2 & 0.05 & -0.05 & 3 & 10 & 8 & 4 \\
& & & & & &  & & & $( 0, \pm 0.2751)$, $ (\pm 3.304,0)$ \\

$0.25\mathrm{e}{-5}$ & 1 & 1 & 2 & 0.05 & -0.05 & 3 & 6 & 8 & 4 \\
& & & & & &  & & & $( 0, \pm 0.1641)$, $ (\pm 3.310,0)$ \\

$0.25\mathrm{e}{-5}$ & 1 & 1 & 2 & 0.5 & -0.005 & 3 & 10 & 8 & 2 \\
& & & & & &  & & &  $ (\pm 33.901,0)$ \\

$0.25\mathrm{e}{-5}$ & 1 & 1 & 2 & 0.6 & 0.5 & 3 & 10 & 8 & 4 \\
& & & & & &  & & & $( 0, \pm 0.1798)$, $ (\pm 15.008,0)$ \\

$0.25\mathrm{e}{-5}$ & 1 & 1 & 2 & 0.5& 0.6 & 3 & 10 & 12 & 4 \\
& & & & & &  & & & $( \pm 13.537, \pm 4.583)$ \\

1 & $1\mathrm{e}{-7}$ & $1\mathrm{e}{-7}$ & $2\mathrm{e}{-7}$ & $1\mathrm{e}{-2}$ & $1.96\mathrm{e}{-2}$ & 3& 10 & 12 & 6 \\
& & & & & &  & & & $( \pm 49.837, \pm 7.099)$,$(\pm 0.2552, 0)$ \\

1 & $1\mathrm{e}{-7}$ & $1\mathrm{e}{-7}$ & $2\mathrm{e}{-7}$ & $1\mathrm{e}{-2}$ & $1.96\mathrm{e}{-2}$ & 3& 5 & 12 & 6 \\
& & & & & &  & & & $( \pm 49.837, \pm 7.099)$,$(\pm 0.1276, 0)$ \\

1 & $1\mathrm{e}{-7}$ & $1\mathrm{e}{-7}$ & $2\mathrm{e}{-7}$ & $1\mathrm{e}{-2}$ & $1.96\mathrm{e}{-2}$ & 3& 14 & 12 & 6 \\
& & & & & &  & & & $( \pm 49.837, \pm 7.099)$,$(\pm 0.2552, 0)$ \\

1 & $1\mathrm{e}{-7}$ & $1\mathrm{e}{-7}$ & $2\mathrm{e}{-7}$ & $1\mathrm{e}{-2}$ & $1.96\mathrm{e}{-2}$ & 3& 0.01 & 8 & 2 \\
& & & & & &  & & & $(0 \pm 7.057)$ \\

2 & $1\mathrm{e}{-7}$ & $1\mathrm{e}{-7}$ & $2\mathrm{e}{-7}$ & $1\mathrm{e}{-2}$ & $1.96\mathrm{e}{-2}$ & 3& 10 & 12 & 6 \\
& & & & & &  & & & $( \pm 35.240, \pm 5.020)$,$(\pm 0.2552, 0)$ \\

1 & $1\mathrm{e}{-7}$ & $1\mathrm{e}{-7}$ & $2\mathrm{e}{-7}$ & $1\mathrm{e}{-1}$ & $1.96\mathrm{e}{-1}$ & 3& 10 & 12 & 6 \\
& & & & & &  & & & $( \pm 50.027, \pm 22.450)$,$(\pm 0.2552, 0)$ \\

1 & $1\mathrm{e}{-6}$ & $1\mathrm{e}{-6}$ & $2\mathrm{e}{-6}$ & $1\mathrm{e}{-2}$ & $1.96\mathrm{e}{-2}$ & 3 & 10 & 12 & 6 \\
& & & & & &  & & & $( \pm 15.758, \pm 2.245)$,$(\pm 0.2553, 0)$ \\

1 & $1\mathrm{e}{-6}$ & $1\mathrm{e}{-6}$ & $2\mathrm{e}{-6}$ & $1\mathrm{e}{-2}$ & $1.96\mathrm{e}{-2}$ & 3 & 5 & 12 & 6 \\
& & & & & &  & & & $( \pm 15.760, \pm 2.245)$,$(\pm 0.1276, 0)$ \\

1 & $1\mathrm{e}{-4}$ & $1\mathrm{e}{-4}$ & $2\mathrm{e}{-4}$ & $1\mathrm{e}{-2}$ & $1.96\mathrm{e}{-2}$ & 3 & 5 & 12 & 6 \\
& & & & & &  & & & $( \pm 1.571, \pm 0.2245)$,$(\pm 0.1280, 0)$ \\
0.25 & $1\mathrm{e}{-7}$ & 0 & 0 & $0.5\mathrm{e}{-2}$ & $-0.5\mathrm{e}{-2}$ & 3 & 10 & 6 & 2 \\
& & & & & &  & & & $(0, \pm 0.2592)$\\

0.25 & $1\mathrm{e}{-7}$ & 0 & 0 & $0.5\mathrm{e}{-2}$ & $-0.5\mathrm{e}{-2}$ & 3 & 5 & 6 & 2 \\
& & & & & &  & & & $(0, \pm 0.1296)$\\

0.25 & $1\mathrm{e}{-7}$ & 0 & 0 & $0.5\mathrm{e}{-2}$ & $-0.5\mathrm{e}{-2}$ & 0.01 & 10 & 4 & 2 \\
& & & & & &  & & & $(\pm 9.918, 0)$\\

0.25 & $1\mathrm{e}{-7}$ & 0 & 0 & $0.5\mathrm{e}{-2}$ & $-0.5\mathrm{e}{-2}$ & 1.5 & 10 & 4 & 2 \\
& & & & & &  & & & $(0, \pm 0.5270)$\\

$0.25\mathrm{e}{-1}$ & $1\mathrm{e}{-7}$ & 0 & 0 & $0.5\mathrm{e}{-2}$ & $-0.5\mathrm{e}{-2}$ & 3 & 10 & 6 & 2 \\
& & & & & &  & & & $(0, \pm 0.2592)$\\

$0.25\mathrm{e}{+1}$ & $1\mathrm{e}{-7}$ & 0 & 0 & $0.5\mathrm{e}{-2}$ & $-0.5\mathrm{e}{-2}$ & 3 & 10 & 6 & 2 \\
& & & & & &  & & & $(0, \pm 0.2592)$\\

0.25 & $1\mathrm{e}{-7}$ & 0 & 0 & 0.5 & -0.5 & 3 & 10 & 4 & 2 \\
& & & & & &  & & & $(\pm 142.901, 0)$\\

0.25 & $1\mathrm{e}{-5}$ & 1 & 0 & $0.5\mathrm{e}{-2}$ & $-0.5\mathrm{e}{-2}$ & 3 & 10 & 6 & 2 \\
& & & & & &  & & & $(\pm 10.011, 0)$\\

0.25 & $1\mathrm{e}{-5}$ & 0 & 1 & $0.5\mathrm{e}{-2}$ & $-0.5\mathrm{e}{-2}$ & 3 & 10 & 8 & 4 \\
& & & & & &  & & & $(\pm 10.011, \pm 0.0313)$\\

$1\mathrm{e}{-3}$ & 1 & 1 & 2 & 1 & 1.96 & 3 & 10 & 12 & 4 \\
& & & & & &  & & & $(\pm 0.4972, \pm 0.7099)$ \\

$1\mathrm{e}{-3}$ & 1 & 1 & 2 & 1 & 1.96 & 3 & 1.4 & 12 & 4 \\
& & & & & &  & & & $(\pm 0.5184, \pm 0.7099)$ \\

$1\mathrm{e}{-3}$ & 10 & 10 & 20 & 1 & 1.96 & 3 & 10 & 4 & 2 \\
& & & & & &  & & & $(0, \pm 0.2141)$\\

$1\mathrm{e}{-3}$ & 100 & 100 & 200 & 1 & 1.96 & 3 & 10 & 0 & 0 \\

$1\mathrm{e}{-3}$ & 100 & 100 & 200 & 1 & 1.96 & 3 & 1.4 & 12 & 4 \\
& & & & & &  & & & $(\pm 0.0474, \pm 0.0710)$\\

$1\mathrm{e}{-3}$ & 100 & 100 & 200 & 1 & 1.96 & 3.09 & 1.4 & 12 & 4 \\ & & & & & &  & & & $(\pm 0.0491, \pm 0.0710)$\\
$0.25\mathrm{e}{-5}$ & 1 & 1 & 2 & 1& 1 & 3 & 10 & 8 & 0 \\

$0.25\mathrm{e}{-4}$ & 1 & 1 & 2 & $0.05\mathrm{e}{+1}$ & $-0.05\mathrm{e}{+1}$ & 3.09 & 10 & 4 & 2 \\
& & & & & &  & & & $( \pm 4.583, 0)$ \\
$0.25\mathrm{e}{-5}$ & 1000 & 1000 & 2000 & 0.05& -0.05 & 3 & 10 & 0 & 0 \\

$0.25\mathrm{e}{-5}$ & 1000 & 1000 & 2000 & 0.05 & -0.05 & 3 & 1.4 & 8 & 4 \\ & & & & & &  & & & $( 0, \pm 0.0388)$,$(\pm 0.3295, 0)$ \\

$0.25\mathrm{e}{-5}$ & 1000 & 1000 & 2000 & 0.05 & -0.05 & 3 & 0.01 & 8 & 4 \\ & & & & & &  & & & $( 0, \pm 0.0003)$,$(\pm 0.3313, 0)$ \\

$1\mathrm{e}{-2}$ & 1 & 1 & 2 & 1 & 1.96 & 3 & 10 & 4 & 2 \\
& & & & & &  & & & $(0, \pm 0.2142)$ \\

$1\mathrm{e}{-2}$ & 1 & 1 & 2 & 1 & 1.96 & 3 & 14 & 0 & 0 \\

$1\mathrm{e}{-2}$ & 1 & 1 & 2 & 1 & 1.96 & 3 & 1.4 & 12 & 4 \\
& & & & & &  & & & $(\pm 0.1627, \pm 0.2242)$ \\

$1\mathrm{e}{-2}$ & 1 & 1 & 2 & 1 & 1.96 & 3 & 5 & 12 & 4 \\
& & & & & &  & & & $(\pm 0.1452, \pm 0.2242)$ \\

$1\mathrm{e}{-2}$ & 1 & 1 & 2 & 1 & 1.96 & 0.01 & 10 & 0 & 0 \\

\hline 
\end{longtable}
\end{footnotesize}
\noindent
When $ \lambda $ = 0.25 $\times 10^{-5}$, $a_{x x} =a_{x y} = 1  $, $ a_{y y} = 2$, $ Z^2_{x} = 0.05$ and $ Z^2_{y} = -0.05$ then in presence of both magnetic field and rotation (~Table \ref{TAble_potential_magnetic_rotation}), the number of stationary points as well as the number of stable centers increase. As B increases the distance between the stable centers on the Y-axis increases and the distance between the stable centers on the X-axis slightly decreases. On the other hand, the effect of rotation is opposite to that magnetic field. When $\Omega$ increases the stable centers on both axes come closer to each other. The depth of the DWP is small (\citealt{Mondal_2021}) for this set of parameter values. Here, a combination of two binary pairs and two field stars form at higher values of both B and $ \Omega$. But, the distance between the field stars is higher compared to the previous case i.e., in the presence of only magnetic field. In the presence of very low rotation, there are no significant differences from the case of only magnetic fields. Again when B is high ($ \sim 10$ $\mu G$) and $ \Omega$ is small ($ \sim 0.01$ \,km\,s$^{-1}$\,pc$^{-1}$ ) then the formation of field stars is more likely than binary pairs or multiple systems. When $ \lambda = 1$, $a_{x x} =a_{x y} = 10^{-7} $, $ a_{y y} = 2\times 10^{-7}$, $ Z^2_{x} = 10^{-2}$ and $ Z^2_{y} = 1.966\times 10^{-2}$, then in the presence of a strong magnetic field and strong rotation formation of a combination of four field stars and two binary pairs is possible.  These types of scenarios may be seen in the filaments having less or moderate gravitational attraction.

\noindent 
In a moderately dense FMC ($ \lambda = 10^{-3}$, $a_{x x} =a_{x y} = 1 $, $ a_{y y} = 2$, $ Z^2_{x} = 1$ and $ Z^2_{y} = 1.966$), the presence of strong rotation and a strong magnetic field increase the distances between the stable centers and four field stars are formed. On the other hand, the presence of strong rotation with low magnetic field forces to form a combination of two binary pairs and two field stars. Binary star formation is more frequent than field stars in the  presence of both a strong magnetic field and strong rotation in a moderately high dense ($ \lambda = 10^{-3}$, $a_{x x} =a_{x y} = 10 $, $ a_{y y} = 20$, $ Z^2_{x} = 1$ and $ Z^2_{y} = 1.966$) FMC.

\noindent
When $ \lambda = 10^{-2}$, $a_{x x} =a_{x y} = 1  $, $ a_{y y} = 2$, $ Z^2_{x} = 1$ and $ Z^2_{y} = 1.96$, then the depth of the potential is higher compared to the depth for the previous set of parameters. In this case, the effect of the magnetic field is more pronounced than rotation. Here, in presence of strong rotation with a low magnetic field binary pairs are formed. For strong rotation and lower (B $ \sim 1.4 $ $\mu G$) or moderate (B$ \sim 5 $ $\mu G$) values of B  stars are formed in stellar associations, whereas, when B is high enough ($ \sim 10$ $\mu G$ or more) there is no stable center irrespective of the strength of rotation, that means when the depth of the DWP is greater i.e for dense filaments if the magnetic field is high then the dynamical system of molecular clouds would diverse rather than converge around some stable centers. 

\section{Conclusions}
\label{sec:conclusions} 
This study employs a synthetic model that leverages the double well potential concept to forecast various star formation scenarios within different categories of rotating FMCs under the influence of a magnetic field.\\
(i) When the depth of the potential is low, i.e., for lesser-dense FMC most of the stars are formed as field stars in the presence of weak as well as strong magnetic fields. The effect of the magnetic field on the number of stars is less. On the other hand, rotation acts opposite to that magnetic field. When clouds rotate fast the distance between the stable centers decreases. The presence of a strong rotation in a strongly magnetized FMC Plays a crucial role in forming a combination of two binary pairs and two field stars.
(ii) When the FMCs have an intermediate density distribution, then in the presence of a strong magnetic field  binary stars form, whereas, in the presence of a low magnetic field stellar association may form. But, from strongly magnetized and strongly rotating moderately dense clouds field stars are formed. In a weakly magnetized cloud, the presence of strong rotation influences to form a combination of two binary pairs and two field stars.
(iii) For moderately high or very high dense FMC, a low magnetic field may help to form binary stars or stellar associations. But, a strong magnetic field strongly acts against star formation, and no star forms. In this case, the effect of magnetic field is more pronounced than rotation. In a rapidly rotating and weakly magnetized cloud binary stars are formed, whereas, strong rotation and moderate magnetic field help to form stellar associations.

\section{Appendix }
\label{appendix}
\subsection{Scaling}
\noindent
$ \ddot X = 2 Z^2_{x}X -4\lambda(a_{xx}X^3+a_{xy}X Y^2)-\frac{1}{\rho}\frac{B^2 X}{X^2+Y^2} +\Omega^2 X +2\Omega \dot Y.$
 Let us put,$ X=x \times 10^{17}\,cm$, $ Y= y \times 10^{17}\,cm$, and $ T = t \times 10^{13}\,s$.  Then the above equation reduces to
 $ \frac{10^{17}}{10^{26}} \ddot x = 2 Z^2_{x}(x \times 10^{17}) -4\lambda(a_{xx}x^3+a_{x y}x y^2) \times 10^{51}-\frac{1}{\rho}\frac{B^2 x}{x^2+y^2} \frac{1}{10^{17}} +\Omega^2 x \times 10^{17} +2\Omega \dot y \times \frac{10^{17}}{10^{13}}.$
 Or, $ \ddot x = 2 (Z^2_{x} \times 10^{26})x -4( \lambda \times 10^{60})(a_{xx}x^3+4a_{x y}x y^2) -\frac{1}{\rho}\frac{B^2 x}{x^2+y^2}10^{-8} +\Omega^2 x \times 10^{26} +2\Omega \dot y \times 10^{13}.$
 Or, $ \ddot x = 2 Z^{'2}_{x}x -4\lambda'(a_{xx}x^3+a_{x y}x y^2)-(\frac{B^{2} \times 10^{-8}}{\rho}) \frac{x}{x^2+y^2} +(\Omega \times 10^{13})^{2} x +2(\Omega \times 10^{13} )\dot y.$
 Where $ Z^{'2}_{x} = Z^2_{x} \times 10^{26} $ and $ \lambda' = \lambda \times 10^{60} $. Now let a, b are two constants such that their ranges are from 1.4 to 14, and from 0.01 to 3 multiplied by 0.3241 respectively.So, I can write $B = a \times 10^{-6} \, G $, and $ \Omega = b \times 10^{-13}$ $s^{-1}$. Also $ \rho = 1.67 \times 10^{-20}$ $ g \, cm^{-3}$. Therefore, $ \frac{B^2}{\rho} \simeq 10^8$, $ \Omega^2 \simeq 10^{-26}$, and $ \Omega = 10^{-13}$. Thus the equation of motion along the x axis in reduce form is 
 $ \ddot x = 2 Z'^2_{x}x -4\lambda'(a_{xx}x^3+a_{x y}x y^2)-\frac{1}{\rho'}\frac{B'^2 x}{x^2+y^2} +\Omega'^2 x +2\Omega' \dot y$,
 Where B'=a, $ \rho' =1.67$, $ \Omega' = b$. Since $Z'^2_{x}$ and $ \lambda'$ are arbitrary constants, the form of this reduce equation is same with original equation before scaling. For the equation of motion along y axis, after scaling I get the same type of equation as original equation with  B'=a, $ \rho' =1.67$, $ \Omega' = b$.
 
=========================================================

\bibliography{sn-bibliography}

\begin{thebibliography}{43}
\providecommand{\natexlab}[1]{#1}
\providecommand{\url}[1]{{#1}}
\providecommand{\urlprefix}{URL }
\providecommand{\doi}[1]{\url{https://doi.org/#1}}
\providecommand{\eprint}[2][]{\url{#2}}
 \bibcommenthead

\bibitem[{Aghili and Kokabi(2017)}]{Aghili2017}
Aghili P, Kokabi K (2017) Astrophys \& Space Sci 362:64

\bibitem[{Bally et~al(1987)Bally, Stark, Wilson, and Henkel}]{Bally1987}
Bally J, Stark A, Wilson R, et~al (1987) ApJS 65:13

\bibitem[{Bally et~al(1988)Bally, Stark, Wilson, and Henkel}]{Bally1988}
Bally J, Stark A, Wilson R, et~al (1988) ApJS 324:223

\bibitem[{Barranco and Goodman(1998)}]{Barranco1998}
Barranco JA, Goodman AA (1998) ApJ 504:207

\bibitem[{{Beck}(2015)}]{Beck_2015}
{Beck} R (2015) {Magnetic fields in spiral galaxies}. The Astronomy and Astrophysics Review 24:4. \doi{10.1007/s00159-015-0084-4}, {\href{https://arxiv.org/abs/1509.04522}{{https://arxiv.org/abs/arXiv:1509.04522}}} {[astro-ph.GA]}

\bibitem[{Boss(2001)}]{Boss_2001}
Boss AP (2001) Impact of magnetic fields on fragmentation. Symposium - International Astronomical Union 200:371–380. \doi{10.1017/S0074180900225424}

\bibitem[{Brainel et~al(2018)Brainel, Rosolowsky, Gratier, Corbelli, and Schuster}]{Brainel2018}
Brainel J, Rosolowsky E, Gratier P, et~al (2018) A\&A 612:A51

\bibitem[{Burkert and Bodenheimer(2000)}]{Burkert2000}
Burkert A, Bodenheimer P (2000) ApJ 543:822--830

\bibitem[{Carlberg and Pudritz(1990)}]{Carlberg1990}
Carlberg RG, Pudritz RE (1990) MNRAS 247:353--366

\bibitem[{Caselli et~al(1999)Caselli, Walmsley, Tafalla, Dore, and Myers}]{Caselli1999}
Caselli P, Walmsley C, Tafalla M, et~al (1999) ApJ 523:165

\bibitem[{Fiege and Pudritz(2000)}]{Fiege2000}
Fiege JD, Pudritz RE (2000) MNRAS 311:85

\bibitem[{Flagey et~al(2009)Flagey, Noriega-Crespo, and Boulanger}]{Flagey2009}
Flagey N, Noriega-Crespo A, Boulanger F (2009) ApJ 701:1450

\bibitem[{Gammie and Ostriker(1996)}]{Gammie1996}
Gammie C, Ostriker E (1996) ApJ 466:814--830

\bibitem[{Goldreich and Kwan(1974)}]{Goldreich1974}
Goldreich P, Kwan J (1974) ApJ 189:441--453

\bibitem[{Goldsmith and Langer(1978)}]{Goldsmith1978}
Goldsmith P, Langer W (1978) ApJ 222:881

\bibitem[{Goodman et~al(1993)Goodman, Benson, Fuller, and Myers}]{Goodman1993}
Goodman A, Benson P, Fuller G, et~al (1993) ApJ 406:528

\bibitem[{Hartmann(2002)}]{Hartmann2002}
Hartmann L (2002) ApJ 578:914

\bibitem[{{Heiles} et~al(1993){Heiles}, {Goodman}, {McKee}, and {Zweibel}}]{Heiles1993}
{Heiles} C, {Goodman} AA, {McKee} CF, et~al (1993) {Magnetic Fields in Star-Forming Regions - Observations}. In: {Levy} EH, {Lunine} JI (eds) Protostars and Planets III, p 279

\bibitem[{Hennebelle and Inutsuka(2019)}]{Hennebelle2019}
Hennebelle P, Inutsuka Si (2019) The role of magnetic field in molecular cloud formation and evolution. Front Astron Space Sci 6:5

\bibitem[{Herbst and Klemperer(1973)}]{Herbst1973}
Herbst E, Klemperer W (1973) ApJ 185:505

\bibitem[{Inutsuka(2018)}]{inutsuka_2018}
Inutsuka Si (2018) The role of magnetic field in the formation and evolution of filamentary molecular clouds. Proceedings of the International Astronomical Union 14(A30):100–100. \doi{10.1017/S1743921319003557}

\bibitem[{Jiménez-Esteban et~al(2019)Jiménez-Esteban, Solano, and Rodrigo}]{Jiménez-Esteban_2019}
Jiménez-Esteban FM, Solano E, Rodrigo C (2019) A catalog of wide binary and multiple systems of bright stars from gaia-dr2 and the virtual observatory. The Astronomical Journal 157(2):78. \doi{10.3847/1538-3881/aafacc}, \urlprefix\url{https://dx.doi.org/10.3847/1538-3881/aafacc}

\bibitem[{Kainulainen et~al(2016)Kainulainen, Hacar, Alves2, Beuther, Bouy, and Tafalla}]{Kainulainen2016}
Kainulainen J, Hacar A, Alves2 J, et~al (2016) A\&A 586:A27

\bibitem[{Khesali et~al(2014)Khesali, Kokabil, Faghei2, and Nejad-Asghar}]{Khesali2014}
Khesali A, Kokabil K, Faghei2 K, et~al (2014) Evolution of filamentary molecular clouds in the presence of magnetic fields. RAA 14:66--76

\bibitem[{Klassen et~al(2016)Klassen, Pudritz, and Kirk}]{Klassen2016}
Klassen M, Pudritz RE, Kirk H (2016) {Filamentary flow and magnetic geometry in evolving cluster-forming molecular cloud clumps}. MNRAS 465(2):2254--2276

\bibitem[{{Krumholz} and {Federrath}(2019)}]{Krumholz_2019}
{Krumholz} MR, {Federrath} C (2019) {The Role of Magnetic Fields in Setting the Star Formation Rate and the Initial Mass Function}. Frontiers in Astronomy and Space Sciences 6:7. \doi{10.3389/fspas.2019.00007}, {\href{https://arxiv.org/abs/1902.02557}{{https://arxiv.org/abs/arXiv:1902.02557}}} {[astro-ph.GA]}

\bibitem[{Li and Berkert(2016)}]{Li2016}
Li GX, Berkert A (2016) {Constructing multiscale gravitational energy spectra from molecular cloud surface density PDF – interplay between turbulence and gravity}. MNRAS 461:3027

\bibitem[{{McKee} et~al(1993){McKee}, {Zweibel}, {Goodman}, and {Heiles}}]{McKee1993}
{McKee} CF, {Zweibel} EG, {Goodman} AA, et~al (1993) {Magnetic Fields in Star-Forming Regions - Theory}. In: {Levy} EH, {Lunine} JI (eds) Protostars and Planets III, p 327

\bibitem[{{Mondal} and {Chattopadhyay}(2019)}]{Mondal_ASHOK_2019}
{Mondal} A, {Chattopadhyay} T (2019) {Fragmentation of molecular cloud in a polytropic medium}. New Astronomy 66:45--51. \doi{10.1016/j.newast.2018.07.008}

\bibitem[{Mondal et~al(2021)Mondal, Chattopadhyay, and Sen}]{Mondal_2021}
Mondal A, Chattopadhyay T, Sen A (2021) A study on the formation of field, binary or multiple stars: a 2d approach through dynamical system. Astrophysics and Space Science 366

\bibitem[{Mondal and Chattopahdyay(2019)}]{Mondal2019}
Mondal D, Chattopahdyay T (2019) Bulgarian Astronomical Journal 31:16

\bibitem[{Myers(2009)}]{Myers2009}
Myers P (2009) ApJ 700:1609

\bibitem[{Myers(2017)}]{Myers2017}
Myers P (2017) ApJ 838:10(13pp)

\bibitem[{Myers and Goodman(1988)}]{Myers1988}
Myers PC, Goodman AA (1988) ApJ 329:392

\bibitem[{{Nakamura} et~al(1997){Nakamura}, {Uehara}, and {Chiba}}]{Nakamura_1997}
{Nakamura} T, {Uehara} H, {Chiba} T (1997) {The Minimum Mass of the First Stars and the Anthropic Pinciple}. Progress of Theoretical Physics 97(1):169--171. \doi{10.1143/PTP.97.169}, {\href{https://arxiv.org/abs/astro-ph/9612113}{{https://arxiv.org/abs/arXiv:astro-ph/9612113}}} {[astro-ph]}

\bibitem[{Ostriker et~al(1999)Ostriker, Gammie, and Stone}]{Ostriker1999}
Ostriker E, Gammie C, Stone J (1999) ApJ 513:259--274

\bibitem[{Ostriker et~al(2001)Ostriker, Stone, and Gammie}]{Ostriker2001}
Ostriker EC, Stone JM, Gammie CF (2001) ApJ 546:980--1005

\bibitem[{{Pudritz} and {Kevlahan}(2013)}]{Pudritz2013}
{Pudritz} RE, {Kevlahan} NKR (2013) {Shock interactions, turbulence and the origin of the stellar mass spectrum}. RSPTA 371(2003):20120,248--20120,248

\bibitem[{{Schleuning}(1998)}]{Schleuning1998}
{Schleuning} DA (1998) {Far-Infrared and Submillimeter Polarization of OMC-1: Evidence for Magnetically Regulated Star Formation}. ApJ 493(2):811--825

\bibitem[{Schneider and Elmegreen(1979)}]{Schneider1979}
Schneider S, Elmegreen B (1979) ApJS 41:87

\bibitem[{Sofue et~al(1986)Sofue, Fujimoto, and Wielebinski}]{Sofue_1986}
Sofue Y, Fujimoto M, Wielebinski R (1986) Global structure of magnetic fields in spiral galaxies. Annual Review of Astronomy and Astrophysics 24(1):459--497. \doi{10.1146/annurev.aa.24.090186.002331}

\bibitem[{Tokuda et~al(2019)Tokuda, Fukui, Harada, and et~al.}]{Tokuda2019}
Tokuda K, Fukui Y, Harada R, et~al (2019) ApJ 886:15

\bibitem[{{Whitworth} and {Stamatellos}(2006)}]{Whitworth_2006}
{Whitworth} AP, {Stamatellos} D (2006) {The minimum mass for star formation, and the origin of binary brown dwarfs}. A \& A 458(3):817--829. \doi{10.1051/0004-6361:20065806}, {\href{https://arxiv.org/abs/astro-ph/0610039}{{https://arxiv.org/abs/arXiv:astro-ph/0610039}}} {[astro-ph]}

\end{thebibliography}

\end{document}